\documentstyle[twoside,fleqn,espcrc2]{article}

\newcommand{\AmS}{{\protect\the\textfont2
  A\kern-.1667em\lower.5ex\hbox{M}\kern-.125emS}}

\title{Calculating the Isgur-Wise Function on the Lattice}

\author{Claude W. Bernard,\address{Physics Department, Washington University,
        St. Louis, MO 63130, USA}
        Yue Shen\address{Physics Department, Boston University, Boston,
        MA 02215, USA}
        \thanks{Speaker at the Lattice '92 conference}
        and
        Amarjit Soni\address{Physics Department, Brookhaven National
        Laboratory, Upton, NY 11973, USA}}

\begin{document}

\begin{abstract}
  We calculate the Isgur-Wise function by measuring the heavy-heavy meson
transition matrix element on the lattice. The standard Wilson action is used
for both the heavy and light quarks. Our first numerical results
are presented.
\end{abstract}

\maketitle

\section{Introduction}
Because of the new flavor and spin symmetries in the heavy quark effective
field theory (HQEFT), the heavy-heavy meson decay
matrix element can be simplified and different decay processes can be related
to each other.
For example, in the case of $B \to D$ decay the number of unknown form
factors can be reduced from two to one,
and we have \cite{Wise}
\begin{equation}
<D_{v^\prime} |{\bar c}\gamma_\nu b |B_{v}> = \sqrt{m_B m_D} C_{cb}
\xi (v^\prime \cdot v) (v + v^\prime)_\nu ~,
\label{eq:BtoD}
\end{equation}
where $v_\nu, v^\prime_\nu$ are the four-velocity and $m_B, m_D$ are
the B and D meson mass, respectively. The constant $C_{cb}$
comes from integrating the full QCD contribution from the heavy quark mass
scale
down to a renormalization scale $\mu \ll m_D$
\begin{equation}
C_{cb} = \left [{\alpha_s(m_D)\over\alpha_s(m_B)}\right]^{6/(33-2N)}
\left [{\alpha_s(m_B)\over\alpha_s(\mu)}\right]^{a(v\cdot v^\prime)}~,
\end{equation}
where $a(v\cdot v^\prime)$ is a slowly varying function of $v\cdot v^\prime$
and vanishes at $v=v^\prime$ \cite{Wise}.
The Isgur-Wise function $\xi (v^\prime \cdot v)$ represents the interactions
of the light degrees of freedom in the heavy meson system and can thus be
calculated only by nonperturbative methods.

On the lattice the heavy meson system can be studied in two different
approaches. One is to keep the heavy quark dynamical by using the standard
Wilson action. This may require extrapolation to the physical heavy meson
mass of interest.
The alternative is to integrate out the heavy
quark first and
derive an effective action including only the light degrees of freedom and
then perform numerical simulation using this effective action. Here we stay
with the first approach.

Using the flavor symmetry of HQEFT the Isgur-Wise function
relevant to the
$B \to D$ decay of Eq. (\ref{eq:BtoD}) can be obtained
also from the $D \to D$ %
elastic scattering matrix element \cite{Wise}
\begin{equation}
<D_{v^\prime} | {\bar c}\gamma_\nu c|D_{v}> = m_D C_{cc}(\mu)
\xi (v\cdot v^\prime) (v + v^\prime)_\nu ~,
\label{eq:matrix}
\end{equation}
where
\begin{equation}
C_{cc}(\mu) = \left[{\alpha_s(m_c) \over \alpha_s(\mu)}\right]
^{ a(v\cdot v^\prime)}~.
\label{eq:cfactor}
\end{equation}
This of course requires that the D meson be sufficiently heavy for the onset
of the heavy quark limit (HQL).
Conventionally a $B \to D$ (here we use B and D as generic names for heavy
pseudoscalar mesons, they do not necessarily represent the physical B and
D mesons) transition matrix can be parametrized as
\begin{equation}
<D_{p^\prime}|V_\nu|B_{p}> = f_+(q^2) (p^\prime + p)_\nu + f_-(q^2)(p -
p^\prime)_\nu~,
\label{eq:form1}
\end{equation}
where $q^2 = (p^\prime-p)^2$ is the momentum transfer between the initial
and final states and $V_\nu$ is a vector current.
It is easy to show that in the elastic scattering case one has $f_-(q^2) = 0$.
Using the relation $v_\nu = p_\nu/m$, Eq. (\ref{eq:form1}) becomes
\begin{equation}
<D_{v^\prime}|V_\nu|D_{v}> = m_D f_+(q^2) (v^\prime + v)_\nu ~.
\label{eq:forwd}
\end{equation}
Comparing this with Eq. (\ref{eq:matrix}) one finds the simple relation
between $f_+$ and $\xi$
\begin{equation}
f_+ = C_{cc} \xi ~.
\label{eq:fplusxi}
\end{equation}
The lattice calculation method for $f_+$ has been well established
\cite{Bernard,ELC}, and thus the result can be easily
used to obtain $\xi$.

\section{Considerations in the Lattice Calculation}
At $\beta = 6.0$ the inverse lattice spacing $a^{-1} \approx 2.0 GeV$.
The HQL becomes valid when the heavy quark $Q$ in a heavy
meson has a mass $m_Q \gg \Lambda_{QCD} \approx 0.2GeV$.
The heaviest mass we can take is order of one or less in lattice units,
beyond which one would expect large lattice-spacing
artifacts. At $\beta=6.0$ this
corresponds to a physical mass in the range of D meson. This can be obtained
by setting the hopping parameter for the heavy quark Q to
\cite{Bernard} $\kappa_Q = 0.118$.
For the light quark we take the hopping parameter
$\kappa_q = 0.152-0.155$ and extrapolate to $\kappa_{q,cr}=0.157$
\cite{Bernard}.

{\it How far can $v\cdot v^\prime$ change on the lattice?}
In our calculations we always have either
the initial or the final particle at rest. Thus
\begin{equation}
v\cdot v^\prime = \left\{ \begin{array}{ll}
                          {E_D / m_D} & \mbox{if $v^\prime = (0,0,0,1)$}\\
                          {E_D^\prime / m_D} & \mbox{if $v= (0,0,0,1)$}
                          \end{array} \right.~,
\end{equation}
where $E_D = \sqrt{ m_D^2 + {\vec p}^2} ~.$

For a spatial lattice size $L = 24$, we have injected momenta
\begin{equation}
{\vec p} = {2\pi \over L} (1,0,0) ~, {2\pi \over L} (1,1,0) ~.
\end{equation}
Since $m_D \approx 1.0$ in lattice unit, one gets $v \cdot v^\prime \approx
1.034$ and $1.066$ respectively. To get larger values
 for $v \cdot v^\prime$
one needs to inject larger lattice momenta which would in turn introduce large
statistical noise in the matrix element calculations. Thus in practice
$v \cdot v^\prime$ can not be much more than $\sim 1.1$.

However, this apparent restriction in the range of $v \cdot v^\prime$ in the
lattice calculations has little physical consequence
as the validity of HQL requires that
the momentum transfer between the initial and final light degrees of
freedom,
$\sim \Lambda_{QCD}^2(v\cdot v^\prime - 1)$,
be $\ll \Lambda_{QCD}^2$\cite{Wise}. This
in turn means that $v\cdot v^\prime$ should be close to one.

\begin{table*}[hbt]
\caption{The Isgur-Wise function. The heavy quark hopping parameter $\kappa_Q$
and the lattice sizes are shown in the table.}
\label{tab:Isgur}
\begin{tabular}{lrrrrrr}
\hline
                 & \multicolumn{2}{l}{.118, $16^3\times 39$}
                 & \multicolumn{2}{l}{.118, $24^3\times 39$}
                 & \multicolumn{2}{l}{.135, $24^3\times 39$} \\
\hline
$v\cdot v^\prime$& $1.0567(8)$ & $1.1103(15)$ & $1.0259(5)$ & $1.0512(10)$
& $1.0543(12)$ & $1.1059(23)$ \\
$\xi$            & $1.00(4)$ & $0.99(11)$ & $0.974(20)$ & $0.940(50)$
& $0.952(20)$ & $0.893(40)$  \\
\hline
\end{tabular}
\end{table*}

{\it Removing the lattice artifacts.}
One of the major concerns in this calculation is the size of the
lattice artifacts. Since the heavy meson mass is
near one in lattice units,
the lattice artifacts could be significant. Ultimately the lattice artifacts
can be brought under control either by comparing data at different $\beta$
values or using the lattice improved actions. However, for simulations at
a given $\beta$ value there are several ways to check
the size of the lattice artifacts.

Eq. (\ref{eq:form1}) and consequently Eq. (\ref{eq:forwd})
require Lorentz invariance to hold.
In Euclidean space the Lorentz transformation becomes a four dimentional
Euclidean rotation. The lattice theory, however, does not have exact
Euclidean rotational invariance for finite lattice spacing $a$.
Thus $f_-$ is not exactly zero. The amplitude of $f_-$ (or $f_-/f_+$) gives
a measure for the violation of the Euclidean invariance on lattice.

We can also estimate the size of the lattice artifacts by checking the
simulation results against known continuum matrix element values at some
special points. For example, when both the initial and final D mesons are at
rest, $v = v^\prime =
(0,0,0,1)$, the continuum matrix element of ${\bar c}\gamma_4 c$ is known
because of the quark flavor current conservation \cite{Wise}
\begin{equation}
<D| {\bar c}\gamma_4 c| D> = 2 m_D ~.
\label{eq:norm}
\end{equation}
At this so-called ``recoil-point" we have $\xi (1) = 1.$
Both Eqs. (\ref{eq:norm}) and (\ref{eq:forwd}) will have $O(a)$ corrections
on lattice. These lattice artifacts may come from different origins.
Part of the $O(a)$ effect can be approximately included by
using a normalization factor
\begin{equation}
<\psi(x){\bar \psi}(0)>_{cont} = 2\kappa u_0 e^{m}<\psi(x){\bar
\psi}(0)>_{latt}~,
\end{equation}
where \cite{Lepage}
\begin{equation}
e^m = 1 +  {1\over u_0}\left({1\over 2\kappa}-{1\over 2\kappa_{cr}}\right) ~,
\end{equation}
with $u_0$ the ``tadpole improvement" factor.

Another correction comes from the use of (nonconserved) local
vector current $V_\nu = {\bar c} \gamma_\nu c$ \cite{karsten}.
This effect can be corrected by introducing a rescaling factor $Z_V^{loc}$
in the vector current. In perturbation theory $Z_V^{loc}$
is calculated to be \cite{Martinelli}
\begin{equation}
Z_V^{loc} = 1 - 27.5{g^2\over 16 \pi^2}.
\label{eq:Z_V}
\end{equation}

In general the size of lattice artifacts will be momentum dependent so the
$O(a)$ correction will be different for Eq. (\ref{eq:norm})
and Eq. (\ref{eq:forwd}).
However, since we are using only small momentum injections we may assume that
the leading $O(a)$ correction does not
depend on $p, p^\prime$ and the Lorentz index $\nu$. We will use
Eq. (\ref{eq:norm}) as the normalization condition for matrix element
$<D_{v^\prime}|V_\nu|D_{v}>$. This way both $exp(m)$ and $Z_V^{loc}$ factors
are taken out automatically (together with any other
 multiplicative $O(a)$ factors).

\section{Numerical results}
We use the standard Wilson action for quarks in the quenched limit.
Both heavy and light quarks are treated as dynamical. We use data at
$\beta=6.0$ on $16^3\times 39$ and $24^3\times 39$ lattices. There are
19 configurations on the $16^3\times 39$ lattice and 8 configurations
on the $24^3\times 39$ lattice. The techniques for measuring the two-point
function and the three-point matrix elements are standard \cite{Bernard}.
The fittings are done for time slices from t=10 to t=15.
To reduce the statistical error, we have
used symmetry properties of the Green functions and averaged over $\pm t$
and $\pm {\vec p}$ directions.

Comparing the measured
$f_0$ value at $q^2=0$ ($f_0(0)=f_+(q^2=0)$) to the known continuum
value
$f_0(0) = 1$, we observed that the lattice artifacts are typically $20\%-40\%$
at $\kappa_Q = 0.118$ and less than $10\%$ at $\kappa_Q=0.135$.
We also measured the ratio $f_-/f_+$.
We find the violation of Euclidean invariance is typically $5\%-15\%$
for $\kappa_Q=0.118$ and $3\%-10\%$ for $\kappa_Q=0.135$.
Note that the reduction of
the lattice artifacts when $\kappa_Q$ is changed from $0.118$ to $0.135$ agrees
with our intuitive expectations. One may try to use the factors $Z_V^{loc}$
and $e^{m}$ to remove part of the $O(a)$ effects. At $\beta=6.0$
we get from perturbative calculation Eq. (\ref{eq:Z_V})
that $Z_V^{loc} \approx 0.7$
(using the shifted effective gauge coupling ${\tilde g}^2 \approx 1.7$
as suggested in ref \cite{Lepage-Mackenzie}). The factor $e^{m}$ is about 2 for
$\kappa_Q = .118$ and 1.5 for $\kappa_Q=.135$. Including both these factors
the corrected $f_0(0)$
becomes 1 within errors. This is in agreement with
other observations \cite{Labrenz} that $Z_V^{loc}$ and $e^{m}$ factors
seem to account for the largest part of the lattice artifacts.

\begin{table*}[hbt]
\caption{Comparison of the lattice calculation for the slope
at $v\cdot v^\prime = 1$ with various model calculations.}
\begin{tabular}{llllll} \hline
Lattice & \cite{rosner} & \cite{neubert} &
\cite{jin}& \cite{block}& \cite{Bjorken,Rafael} \\ \hline
 1.0(8) &  1.6(4)  & 1.4(6) & 1.05(20) & 0.65(15) & $0.25 < \rho^2 < 1.42$
\\ \hline
\end{tabular}
\end{table*}

According to the discussions in Section 2, we define $\xi (v\cdot v^\prime)
= f_+/f_0(0)$ and list the results in Table 1. We emphasize that this
definition
removes all momentum-independent
$O(a)$ effects simultaneously, including both
$e^m$ and $Z_V^{loc}$ factors. Note that in Eq. (\ref{eq:fplusxi}) there
is a factor $C_{cc}$ in the connection between $f_+$ and $\xi$.
This factor comes from integrating out the QCD effects from the heavy quark
scale down to a light scale $\mu$. For the lattice calculation, however,
$\mu$ is taken to be $O(1/a)$. Thus in our case $\mu \sim m_D
\sim m_c$. Also $v\cdot v^\prime$ is very close to one. So for practical
purposes we can set $C_{cc} = 1$ according to Eq. (\ref{eq:cfactor}).

We plot our results for the Isgur-Wise function in Fig. 1. For comparison
we also plotted the theoretical bounds on the Isgur-Wise function. The top
and bottom curves are the upper and lower bounds derived in ref. \cite{Rafael}.
They are obtained using the dispersion relation for the two-point
functions, with the requirements of unitarity and causality and with
some assumptions on the
analytic properties of the form factors. The curve in the middle is
an upper bound on the Isgur-Wise function derived from current-algebraic sum
rules \cite{Bjorken}. This is a tighter upper bound.
Our data obtained on the $24^3\times 39$ lattice appear to be, within errors,
inside of the upper bound of Bjorken \cite{Bjorken} and the lower bound of
de~Rafael and Taron\cite{Rafael}.
The data from the $16^3\times 39$ lattice is consistent with the Bjorken
upper bound within rather large errors.

\begin{figure}[htb]
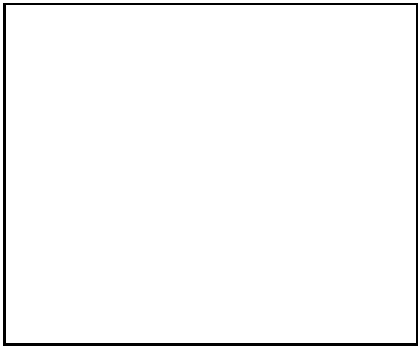

\framebox[55mm]{\rule[-21mm]{0mm}{43mm}}
\caption{The Isgur-Wise function is plotted against $v\cdot
v^\prime$. The open circles represent data on the $16^3\times 39$ lattice.
The solid circles and squares represent data on the $24^3\times 39$ lattice.
The heavy quark hopping parameter is set at $\kappa_Q = 0.118$ (open and
solid circles) and at $\kappa=0.135$ (squares) while the light quark
hopping parameters are extrapolated to the chiral limit $\kappa = \kappa_c
\approx 0.157$.}
\label{fig:figure1}
\end{figure}

Close to $v \cdot v^\prime = 1$ the Isgur-Wise function can be parametrized
as
\begin{equation}
\xi (v\cdot v^\prime) = 1 - \rho^2 (v\cdot v^\prime - 1)
+ O((v\cdot v^\prime-1)^2) ~.
\end{equation}
If we calculate the slope using the data point closest to the $v\cdot v^\prime
=1$ axis, we get $\rho^2 = 1.0(8)$. In Table 2 we list our result along with
$\rho^2$ values estimated by other authors.

Note that the lattice meson mass we used is in the range of physical D meson.
Thus the $O(1/m_Q)$ correction
may be quite significant. For a reliable
calculation of the Isgur-Wise function we need to
repeat this calculation for several different masses and then extrapolate to
the infinite mass limit. For a check on the residual $O(a)$ effects
we plan to repeat our calculation at $\beta=6.3$.

\vspace{0.2cm}
\noindent{\it Acknowledgement}

C.B. was partially supported by the DOE under grant number
DE2FG02-91ER40628.
Y.S. was supported in part under DOE contract DE-FG02-91ER40676 and NSF
contract PHY-9057173, and by funds from the Texas National Research Laboratory
Commission under grant RGFY92B6.
A.S. was supported in part by the DOE under grant
number DE-AC0276CH00016.

The computing for this project was done at the National
Energy Research Supercomputer Center in part under the
``Grand Challenge'' program and at the San Diego Supercomputer Center.


\begin{thebibliography}{9}

\bibitem{Wise} For reviews, see M.~Wise, CALT-68-1721,
published in the Proceedings of the Lake Louise Winter
Institute, 1991 p.222;
H.~Georgi, preprint HUPT-91-A039, 1991.

\bibitem{Lepage} A.~S.~Kronfeld and P.~B.~Mackenzie, private communications;
also see G.~P.~Lepage, Nucl. Phys. B (Proc. Suppl.) 26 (1992) 45.

\bibitem{Bernard} C.~W.~Bernard, A.~X.~El-Khadra and A.~Soni,
Phys. Rev. D43 (1991) 2140.

\bibitem{ELC} M.~Crisafulli, G.~Martinelli, V.~Hill and C.~Sachrajda,
Phys. Lett. B223 (1989) 90.

\bibitem{Labrenz} C.~W.~Bernard, C.~M.~Heard, J.~Labrenz and A.~Soni,
Nucl. Phys. B (Proc. Suppl.) 26 (1992) 385; J. Labrenz, these
proceedings.

\bibitem{karsten} L.~H.~Karsten and J.~H.~Smit, Nucl. Phys. B183 (1981) 103.

\bibitem{Martinelli} G.~Martinelli and Y.~C.~Zhang, Phys. Lett. B123
(1983) 433.

\bibitem{Lepage-Mackenzie} G.~P.~Lepage and P.~B.~Mackenzie,
Nucl. Phys. B (Proc. Suppl.) 20 (1991) 173.

\bibitem{Rafael} E.~de~Rafael and J.~Taron, Phys. Lett. B282
(1992) 215.

\bibitem{Bjorken} J.~D.~Bjorken, SLAC report SLAC-PUB-5278 (1990).

\bibitem{rosner} J.~L.~Rosner, Phys. Rev. D42 (1990) 3732.

\bibitem{neubert} M.~Neubert, Phys. Lett. B264 (1991) 455.

\bibitem{jin} H.~Y.~Jin, C.~S.~Huang and Y.~B.~Dai, AS-ITP-92-37.

\bibitem{block} B.~Block and M.~Shifman, preprint NSF-ITP-92-100,
TPI-MINN-92-32/T, June 1992.

\end{thebibliography}
\end{document}